\documentclass[aps,twocolumn,superscriptaddress]{revtex4}

\usepackage[english]{babel}
\usepackage{graphicx}
\usepackage {subfigure}
\usepackage{xcolor}

\usepackage{amsmath,amssymb,amsfonts,latexsym,fancyhdr,graphicx,times,txfonts}

\usepackage{color}
\usepackage[pdftex,colorlinks=true,allcolors=blue]{hyperref}

\def\bra#1{\mathinner{\langle{#1}|}}
\def\ket#1{\mathinner{|{#1}\rangle}}

\begin{document}

\title{Momentum-resolved spectroscopy of a 1D superfluid using a single atomic impurity}

\author{F. Cosco}
\affiliation{QTF Centre of Excellence, Turku Centre for Quantum Physics, Department of Physics and Astronomy, University of Turku, FI-20014 Turun yliopisto, Finland}
\author{M. Borrelli}
\affiliation{QTF Centre of Excellence, Turku Centre for Quantum Physics, Department of Physics and Astronomy, University of Turku, FI-20014 Turun yliopisto, Finland}
\author{F. Plastina}
\affiliation{Dipartimento di Fisica, Universit\`a della Calabria, 87036, Arcavata di Rende (CS), Italy}
\affiliation{INFN - Gruppo Collegato di Cosenza, Cosenza, Italy}
\author{S. Maniscalco}
\affiliation{QTF Centre of Excellence, Turku Centre for Quantum Physics, Department of Physics and Astronomy, University of Turku, FI-20014 Turun yliopisto, Finland}
\affiliation{QTF Centre of Excellence, Department of Applied Physics, Aalto University School of Science, P.O. Box 11000, FIN-00076 Aalto, Finland}

\begin{abstract}
We present a general and non-invasive probing scheme to perform
full momentum-resolved spectroscopy of a cold atomic gas loaded
into an optical lattice using a single quantum impurity. The
protocol relies on weak collisional interactions and subsequent
population measurements of the impurity. By tuning a few
controllable external parameters the impurity-lattice interaction
can be engineered, and using two sets of measurements, performed
with the impurity in two different positions, the full dispersion
relation of the superfluid phonons can be reliably extracted.
\end{abstract}

\maketitle

\section{Introduction}

Cold atoms in optical lattices allow to engineer and investigate non-trivial Hamiltonian models
in a fully controllable way \cite{bloch1,cazalilla1,bloch2}, with the possibility of tuning the interactions  \cite{primadidi} by
means of Feschbach resonances \cite{dopob2,dopob3}. In these setups  typical condensed matter physics effects and models can
be simulated \cite{vari,andreas,andreas2} with the lack of lattice defects.
In this context, the Bose-Hubbard model is perhaps the most
celebrated example \cite{hubbard,kanamori}. This model has been
extensively studied theoretically
\cite{jaksch1,oosten1,altman1,huber1,barmettler1,walters1} and a
great number of experimental verifications have been performed
\cite{bloch3,stoferle1,paredes1,inguscio1}. Furthermore, recent
experiments in the context of quantum information and simulations
using cold atoms in optical lattices also suggest that the
Bose-Hubbard model can be of practical relevance for technological
applications \cite{endres1}. As for most systems in condensed
matter physics, probing of cold atoms in optical lattices is
usually performed via semi-classical methods that can be rather
invasive or even destructive, depending on the specific technique
or quantity to be measured. 
A prominent example are the superfluid
excitations of a Bose-Hubbard gas. These have been resolved in
energy using techniques such as magnetic gradients \cite{bloch3}
and lattice depth modulation \cite{OPL,OPL2}.
 A full momentum-resolved
spectroscopy leading to the measurement of the superfluid
dispersion relation has been performed using Bragg spectroscopy in
\cite {BRAGG,BRAGG2}. This method, however, relies on two-photon
processes, thus implying the exchange of energy and momentum with
the atoms in the lattice. Although all of these techniques have
been successfully applied, they strongly interfere with the
dynamics of the gas and one may wonder whether it could be
possible to extract similar information without disturbing the
system so much. Very recently, single and controllable quantum
objects have been proposed as an alternative tool to investigate
collective properties of large many-body systems. Successful
examples in optical lattice systems range from transport
properties \cite{TomiSteve} to temperature estimation
\cite{tarxiv} and measure of quantum correlations \cite {elliott2016,streif2016}. Further instances include probing of cold free and trapped
gases \cite{pinja1,pinja1b,jaksch2,fischer,recati}, spin chains
\cite{pinja2}, Fermi systems \cite{demler,plastina1,cetina2016,schmidt2018,schiro},
Coulomb crystals \cite{massimo1,massimo2} and generically critical systems
\cite{zwick}. Recently, an interesting spectroscopic protocol was
presented to study some energy-resolved spectral features of
atomic gases in optical lattices \cite{giammarchi2}. Here, we take
a step further and propose an experimentally feasible application of the protocol outlined in \cite {cosco2017}. We show how to perform full
momentum-resolved spectroscopy of a cold 1D superfluid gas in an
optical lattice using a single atomic impurity. The impurity is
harmonically trapped in an auxiliary potential well and brought
into contact (and interaction) with the surrounding gas. By
properly controlling the coupling strength and the position of the
impurity, it is possible to engineer a two-stage spectroscopic
protocol that allows for the reconstruction of the dispersion
relation $\omega(k)$ of the quasi-particle excitations of the
atomic gas. We call such an impurity a quantum probe.
\section {Protocol}
\begin{figure}
\begin{center}
\includegraphics[scale=0.35]{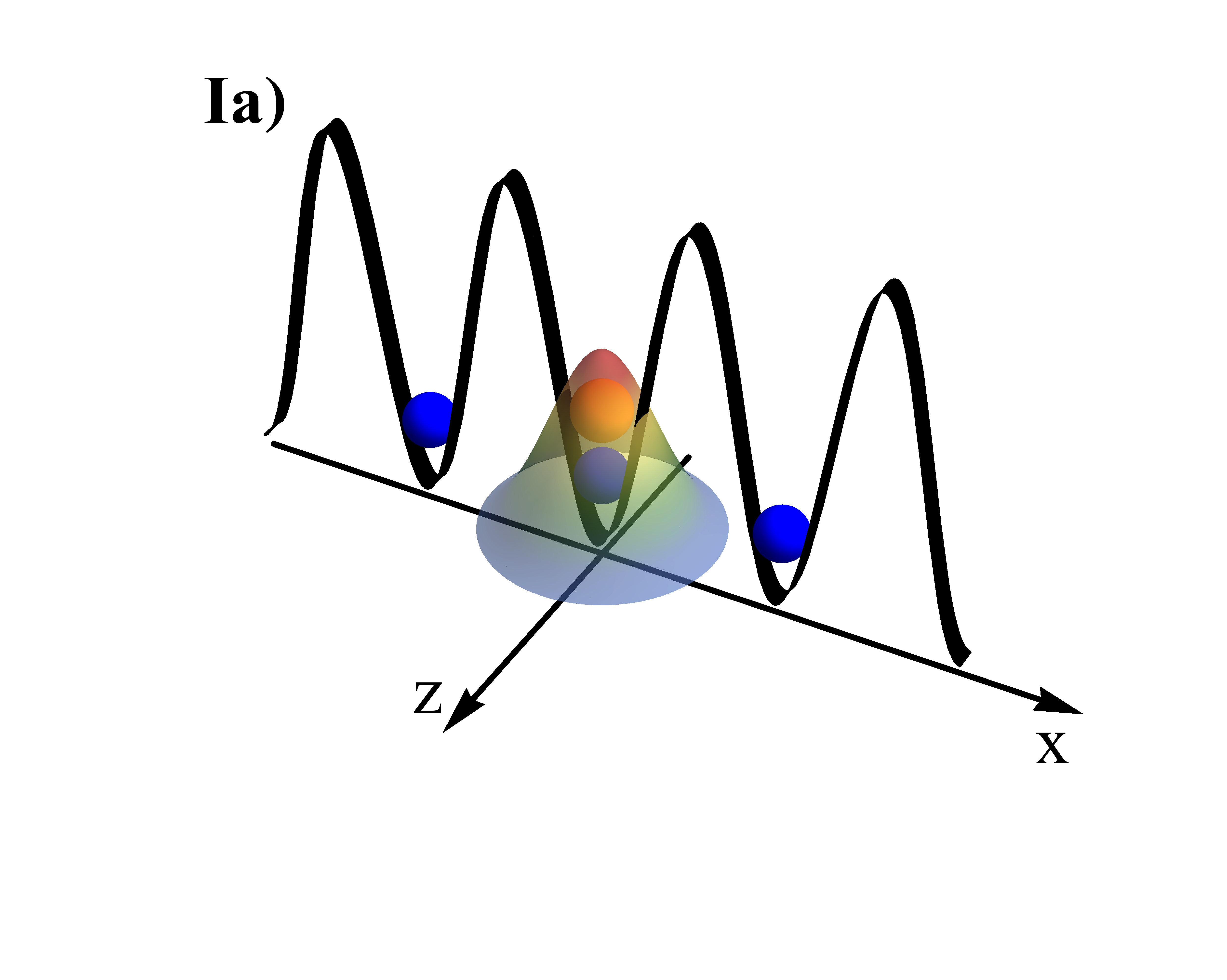}\includegraphics[scale=0.35]{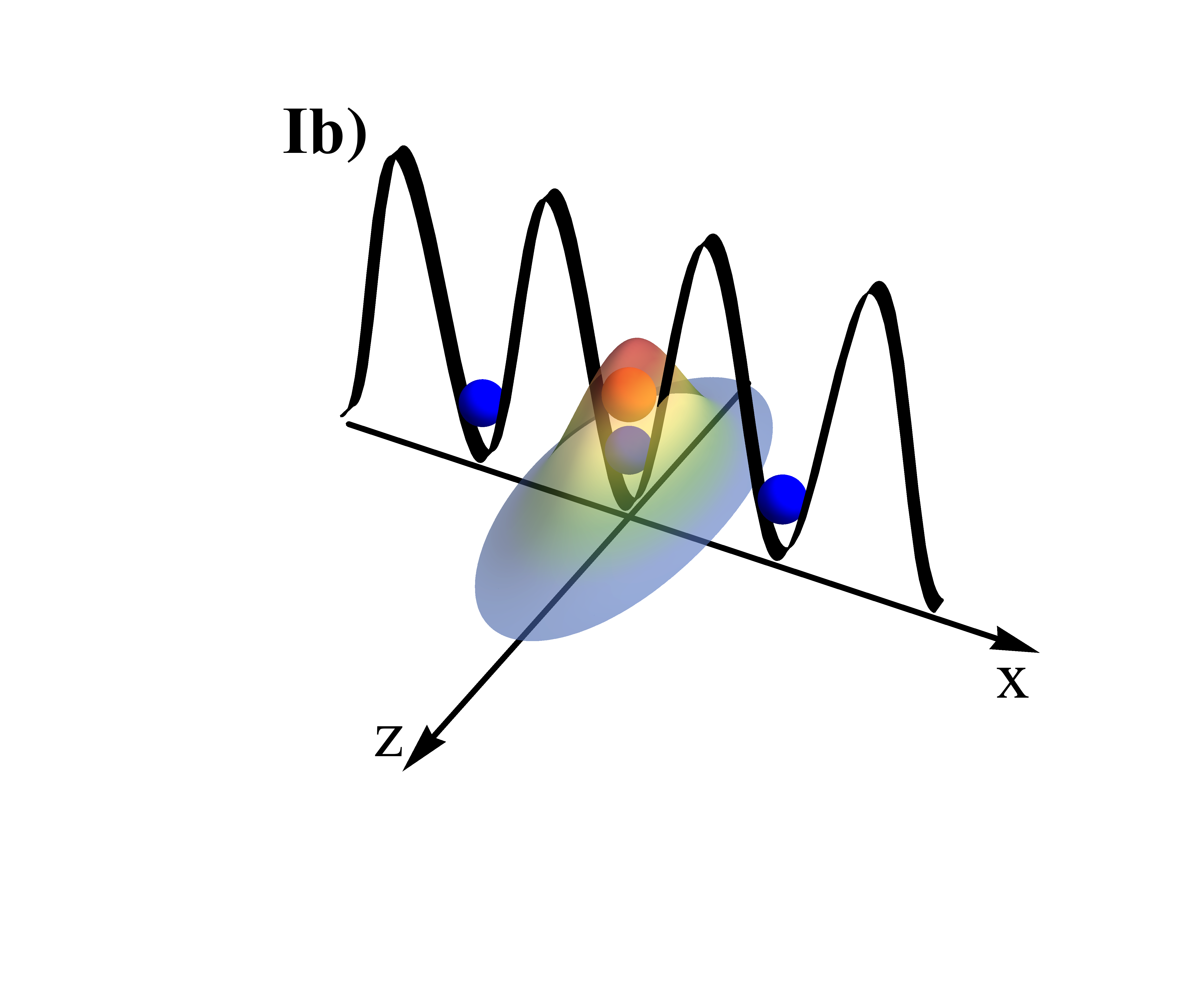}
\includegraphics[scale=0.35]{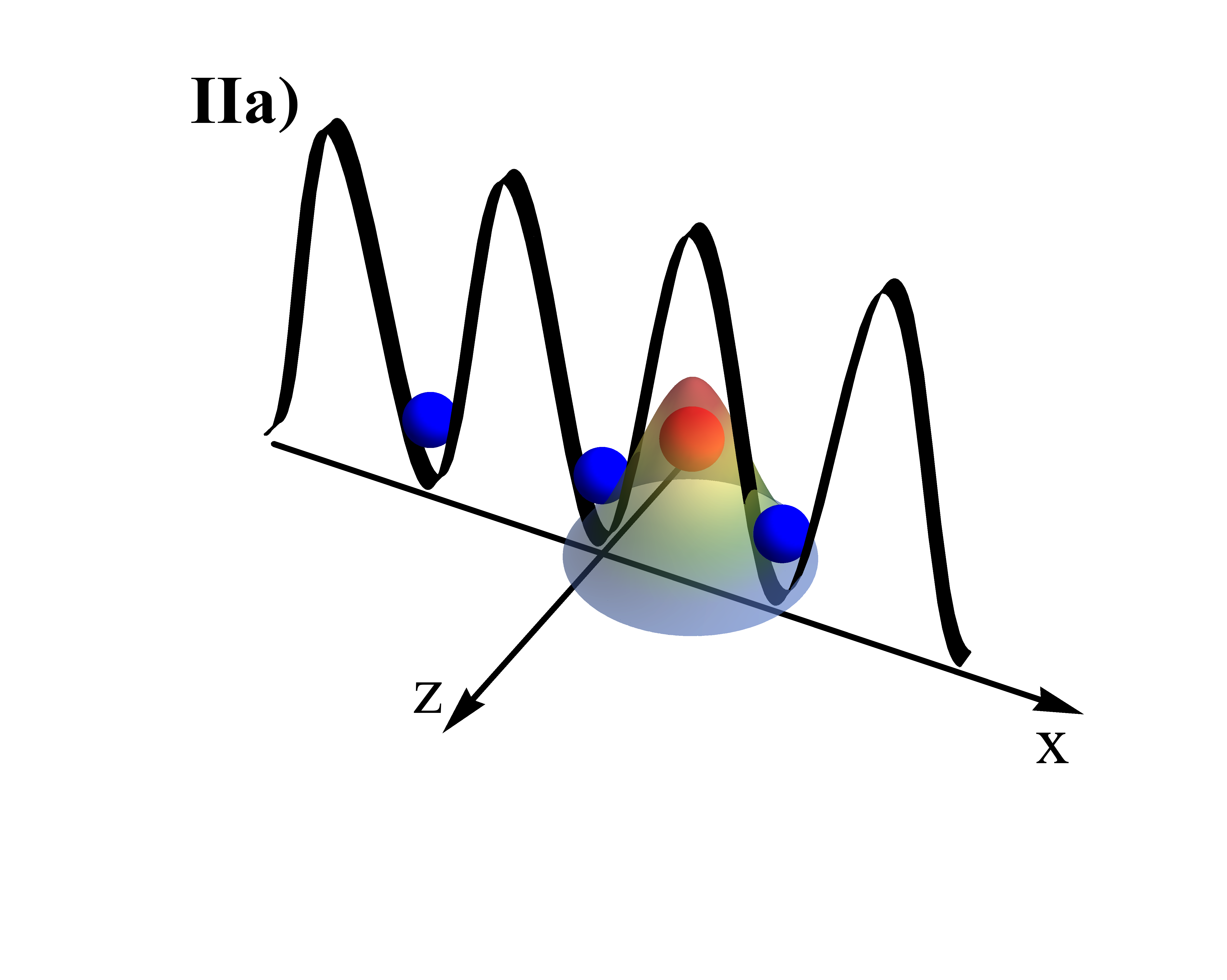}\includegraphics[scale=0.35]{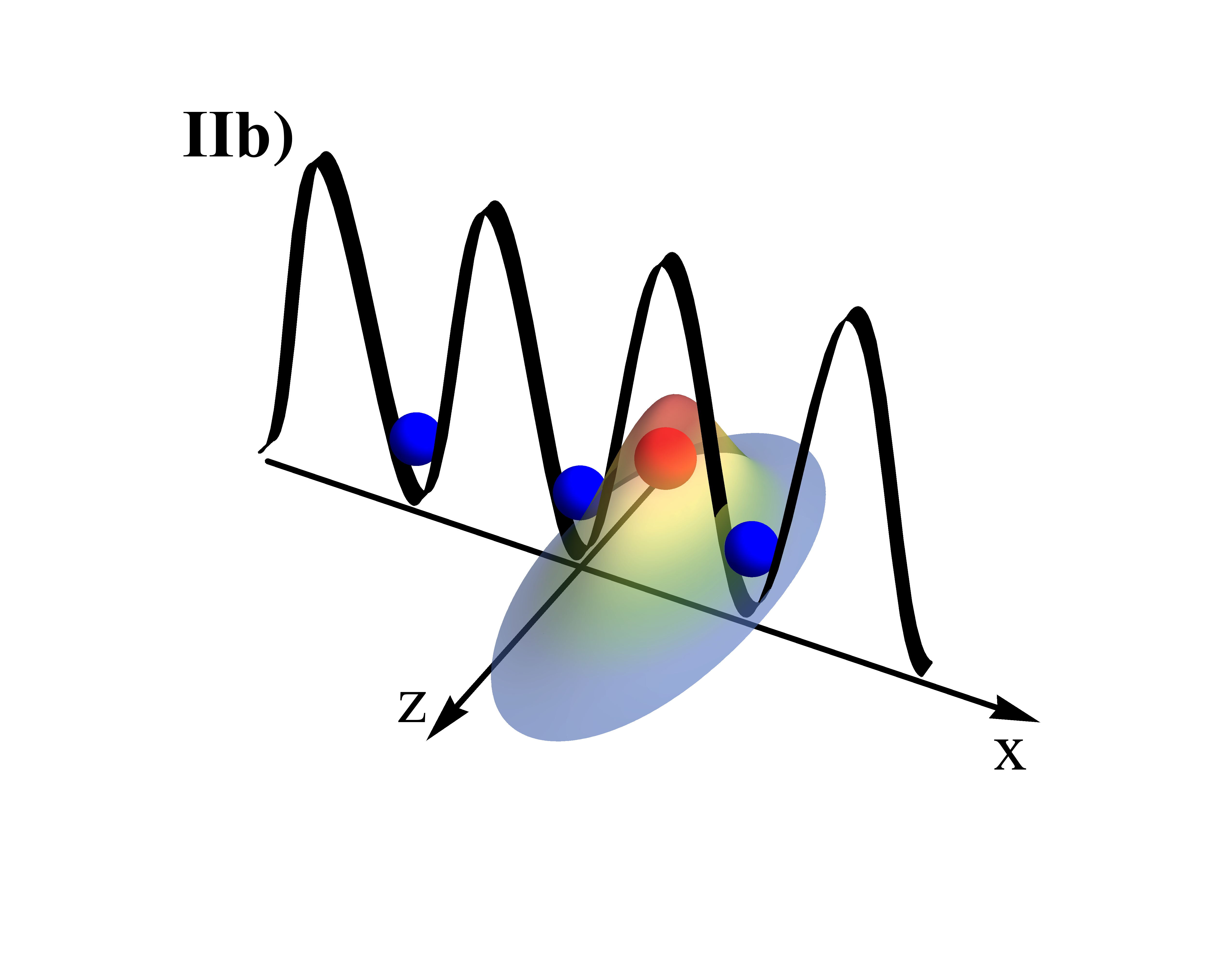}
\caption { Sketch of the two different steps of the protocol. In the upper
panel, the trapped impurity is located at a minimum of the optical
lattice and its ground state wave function  overlaps with the
Wannier state of that site only. The lower panel, instead, shows
the impurity localized near a maximum of the lattice, with a
ground state wave function large enough to couple with both of the
adjacent sites.} \label{scheme}
\end{center}
\end{figure}

Our scheme is depicted in Fig.~\ref{scheme} and
it can be implemented by using a either Rb-K setup \cite{exp1,expref1} or
a modification of the spin-dependent Cs scheme described in \cite{exp2}.
The protocol consists of two subsequent steps in which transitions between energy levels
of the impurity are observed. The transition rates  strongly
depend on the position of the impurity within the lattice. By
combining the outcomes of two sets of energy-resolved measurements
corresponding to different impurity positions, it is possible to
momentum-resolve both the excitation spectrum and spectral density
of the Bose gas. The two key features of our protocol being 1)
measurements are to be performed only on the probe, hence causing
minimal disturbance to the atomic gas, and 2) no momentum exchange
between the probe and the gas ever occurs.
The dynamics of an ensemble of bosonic atoms trapped in a
one-dimensional optical potential and cooled to its lowest energy
band is governed by the Bose-Hubbard Hamiltonian
\cite{hubbard,hubbard2,kanamori}:
\begin{center}
\begin{equation}
\hat H_B = -J \sum_{\langle i,j \rangle} \hat c^\dagger_{i} \hat c_j + \frac {U}{2} \sum_i \hat n_i (\hat n_i -1)-\mu \sum_i \hat n_i,
\label{BHH}
\end{equation}
\end{center}
Here, $\hat {c}^\dagger_{i}, \hat c_i $ are local boson ladder
operators labelled by the lattice site with $\hat n_i\equiv\hat
c^\dagger_{i} \hat c_i$, $\langle \rangle$ in the first sum
selects nearest neighbour sites,  $J$ is the hopping constant, $U$
is the on-site interaction strength and $\mu$ the chemical
potential. Both the static and dynamical properties of the boson
gas described by such a model result from the interplay of two
competing mechanisms: while the hopping between sites tends to
favor the atomic mobility, the positive on-site interaction tends
to localize the particles on the lattice. This results in a
 transition from a superfluid phase ($J\gg U$), in
which atoms hop freely between near sites, to a Mott insulator
phase ($J\ll U$), in which transport is suppressed
\cite{fisher1,bloch3}. In the superfluid phase we can work within
the Bogoliubov approximation and diagonalize Hamiltonian 
(\ref{BHH}) 
in terms of phonon excitations above a uniform
Bose-Einstein condensate. In this regime, the Hamiltonian can be
expressed in terms of phonon-like modes, and reads (from now on,
we set $\hbar=1$) \cite{oosten1} $\hat{H}_B =
\sum_{k}\omega_{k}\hat{b}^\dagger_{k}\hat{b}_k$, in which $\hat
b^\dagger_k, \hat b_k$ are the Bogoliubov ladder operators
describing phonons at energy
$\omega(k)=\sqrt{\epsilon^2_k+2Un_0\epsilon_k}$ with $\epsilon_k =
2J[1-\cos (ka)]$. Here $a$ is the lattice constant, while $n_{0}$
is the density fraction of condensed atoms. We assume that an
atomic impurity trapped in an auxiliary potential well, {\it i.e.}
an atomic quantum dot \cite{recati}, is immersed in the lattice
and interacts with its surrounding atoms. The unperturbed
Hamiltonian of the probe-impurity can be expressed in terms of its
(localized) eigenstates (whose detailed form depends on the shape
of the trapping potential) and reads $\hat H_P = \sum_n \nu_n
\ket{n}\bra{n}$. The coupling to the Bose gas is taken to be of
the density-density type \cite{jaksch2}, with the usual assumption
of contact potential, and reads
\begin{equation}
\begin{split}
  \hat H_{int}=g\sum_{n,m;i,j}  \int dxdydz \; \psi_n^*(x,y,z) \psi_m(x,y,z) & \times \\
 \omega_i^*(x)\omega_j(x)\delta_{i}(y,z)\delta_{j}(y,z)\ket{n}\bra{m}\otimes   \hat c^\dagger_i \hat c_j,
\label{interaction1}
\end{split}
\end{equation}
in which we have assumed a three dimensional spatially extended
probe, although the lattice is effectively one dimensional. Here,
$g$ is the impurity-gas coupling constant, $\psi_m(x,y,z) = \left
\langle x,y,z| m \right \rangle$ is the $m$-th unperturbed
impurity energy eigenfunction, while $\omega_i(x)$ is the Wannier
eigenfunction corresponding to the $i$-th lattice site.
The effective coupling between the impurity and the bosons at the
$i$-th site of the lattice depends upon the overlap integrals $
\varphi_{nm}=\int dxdydz \, \psi^*_n(x,y,z) \psi_m(x,y,z) \omega_i
(x)^2 \delta_{i}(y,z)$. In what follows we assume the probe to be
spatially localized around one specific site that we label
$\mathbf 0$. This allows us to drop the summation over the site
index and simplify Eq.~\ref{interaction1}. Furthermore, by
employing the Bogoliubov approximation and expressing the number
of bosons at site ${\mathbf 0}$ in terms of the Bogoliubov modes
\cite{oosten1}, the interaction Hamiltonian can be rewritten as
\begin{equation}
\hat H_{int}=g\sum_{n,m} \varphi_{nm} \ket{n}\bra{m}\otimes\left[n_0+\sum_k  {\beta_k}{} (\hat b_k^\dagger+\hat b_k )\right],
\label{interaction2}
\end{equation}
in which $\beta_{k}=\sqrt{\frac {n_{0}}{{N_s}}}(u_{k}+v_{k})$,
with $u_{k},v_{k}$ being the Bogoliubov coefficients, whose
analytical expression can be found, e.g., in \cite{oosten1}. While
in real space the impurity couples locally to one specific lattice
site (that is $\mathbf{0}$), in the momentum space it couples to
all of the Bogoliubov modes. The interaction (\ref{interaction2})
describes transitions between different energy levels of the probe
associated to phonon propagating through the lattice. In the
following, we take the probe to be initialized in its unperturbed
ground state $ \ket {0}$, while the gas loaded into the lattice is
in a thermal state $\rho_\beta\propto \exp\left(-\beta\sum_k
\omega(k) \hat b^\dagger_k \hat b_k\right)$. In this way, only
ground-to-excited state transitions of the atomic probe have to be
considered. The probability $\Gamma_{0\rightarrow n} $ for such a
transition to occur within time $t$ can be easily computed in the weak coupling limit

\begin{equation}
\Gamma_{0\rightarrow n}(t)=g^2\varphi_{n0}^2\Bigl\{\Gamma_0
+\sum_k \Gamma^-_k(\omega,t) + \Gamma^+_k(\omega,t) \Bigr \}\, ,
\label{gammafinale1d}
\end{equation}
with $\Gamma_0= \lambda_{1}(\omega_{n},t)n_{0}^{2}$,
$\Gamma^-_k(\omega,t)\equiv \beta_k^2
\lambda_1(\omega+\omega(k),t)(1+n(\omega(k)))$ and
$\Gamma^+_k(\omega,t)\equiv \beta_k^2
\lambda_2(\omega-\omega(k),t)n(\omega(k))$. The latter three
quantities are expressed in terms of the probe transition
frequency $\omega_{n}\equiv \nu_n-\nu_0$, the Bose-Einstein
distribution at temperature $\beta^{-1}$, $n(\omega)$, and two
auxiliary functions $\lambda_{1}(\omega,t)=2\left[1-\cos(\omega t)\right]/\omega^{2},\lambda_2(\omega-\omega(k),t)=\lambda_{1}(\omega-\omega(k),t)$ if $\omega\neq\omega(k)$, $\lambda_2(\omega-\omega(k),t)=t^2$ if $\omega=\omega(k)$.

To go further in the analysis, we consider a specific trapping
potential for the probe and, as a simple and yet physically
relevant example, we analyse the case of an harmonic trap. For the
sake of clarity, we first discuss a  one-dimensional impurity
trap, in which the longitudinal spreading of the impurity wave
function can be neglected. Later on we will move to a more
realistic three dimensional trapping well. 
\subsection {Quantum Probe: Impurity atom in 1D harmonic trap}
In the simple 1-D
harmonic case, the probe eigenenergies are $\nu_n = \nu(n+\frac
{1}{2})$, while the unperturbed eigenfunctions are given in terms
of the Hermite polynomials $H_n$ and read $\psi_n^{(\nu)}(z)=\frac
{( {m \nu})^{1/4}\pi^{-1/4}}{2^{n/2}n!^{1/2}}H_n(\sqrt { {m \nu}}
z) e^{-\frac {m \nu z^2}{2} }$, where $m$ is the impurity mass.
Here the $z$ axis (along which the probe trapping well extends) is
imagined to be orthogonal to the lattice axis; with this spatial
arrangement the interaction Hamiltonian fully satisfies the
localization assumption discussed above. As a side effect of the
harmonic approximation, the parity of the probe eigenstates
implies that transitions are only induced between even numbered
levels, \cite{nota}. Assuming that the minimum of the harmonic
trap coincides with a selected minimum of the optical lattice, the
amplitude $\varphi_{nm}$ entering the probabilities above, becomes
$\varphi_{nm}=\sqrt { {m}} \omega_0^2(0) \sqrt {\nu}
\frac {1}{\pi} (-1)^{n+m} \gamma_n^{1/2} \gamma_m^{1/2}$, in which
$\gamma_n =\frac {\Gamma (n+1/2)}{\Gamma {(n+1)}}$ is the Euler
Gamma function ratio.  As a result, the transition probability
from the ground to the $n$-th excited level reads
\begin{equation}
\Gamma_{0\rightarrow n} = g_n'^2 \nu \lambda_1 (n \nu ,t) n_0^2  +g_n'^2 \nu \sum_k \Gamma^+_k(n\nu,t)+ \Gamma^-_k(n\nu,t),
\end{equation}
in which $g_n'=g\frac { \sqrt { {m}} \omega_0^2(0)}{\pi} \sqrt{
\gamma_n \gamma_0 }$. 
\section {Quantum Probe: Impurity atom in 3D harmonic trap}
In a realistic experimental situation, the
three dimensional spatial extension of the probe wave function has
to be taken into account. We consider a 3D harmonic trap and
assume the trap frequency to be tailored (and controllable) along
one direction orthogonal to the lattice. The confinement in the
two other directions is kept fixed. The unperturbed probe wave
functions are now given by three factors, one for each spatial
coordinate, $\psi_{\bar n} (\mathbf x)= \psi_{n_x}^{(\nu_0)}
(x)\psi_{n_y}^{(\nu_0)} (y)\psi_{n_z}^{(\nu)} (z)$. As before, we
are interested in measuring transition probabilities between
impurity states along the $z$ direction, and assume the $x$ and
$y$ degrees of freedom to be frozen. We therefore need to evaluate
$\Gamma_{\bar 0 \rightarrow (0,0,n_z) }$. As depicted in Fig.
\ref{scheme},  the protocol develops in two subsequent steps with
the probe brought onto {\bf I)} a minimum, and {\bf II)} a maximum
of the optical potential. The information obtained by using these
two steps allows for the reconstruction of the full dispersion
relation $\omega(k)$, that is momentum-resolved spectroscopy of
the gas through measurements on the impurity. As sketched in Fig.
\ref{scheme}, in configuration {\bf I)} the impurity interacts
with one lattice site only, while in configuration {\bf II)}, due
to a suitable choice of the longitudinal confining frequency
$\nu_0$, the impurity is coupled with two adjacent sites at the
same time. The transition probabilities corresponding to the two
positions can be computed as before. For case {\bf I)}, we obtain
an expression which is identical to Eq. (\ref{gammafinale1d}), but
for the pre-factor:
\begin{equation}
\Gamma^{\textrm{\bf I}}_{\bar 0 \rightarrow (0,0,n_z) } = g_{\textrm{\bf I},n}^{2}\nu \Bigg \{\lambda_1 (n \nu ,t) n_0^2 + \sum_k[ \Gamma^+_k(n\nu,t)+ \Gamma^-_k(n\nu,t)]  \Bigg \},
\label{gammaA}
\end{equation}
where the pre-factor  $g_{\textrm{\bf I},n}^2=\frac {g^2}{\nu}
X_{00}^2 Y_{00}^2 Z_{n_z0}^2$ , is expressed in terms of the
spatial overlap $X_{00}=\int dx \, \psi_{n_x=0}^2(x)\omega_0^2(x)$
, and of the constants $Y_{00}=\sqrt { {m}}\frac
{\gamma_0}{\pi}\sqrt {\nu_0}$ and $Z_{n_z0}=(-1)^{n_z}\sqrt {
{m}}\frac {\gamma_{n_z}^{1/2}\gamma_0^{1/2}}{\pi}\sqrt {\nu}$
related to the confinement in the transverse directions. For case
{\bf II)}, assuming that the probe interacts with equal strength
with the two adjacent sites, the transition probability reads
\begin{equation}
\begin{split}
\Gamma^{\textrm{\bf II}}_{\bar 0 \rightarrow (0,0,n_z) } = g_{\textrm{\bf II},n}^2 \nu\Bigg \{ 2\lambda_1 (n \nu ,t) n_0^2   \\
+   \sum_k (1+\cos (ka))[\Gamma^+_k(n\nu,t)+ \Gamma^-_k(n\nu,t)] \Bigg \},
\end{split}
\label{gammaB}
\end{equation}
where the new pre-factor, $g_{\textrm{\bf II},n}^2=2\frac {g^2}{\nu}  X_{00}^2
Y_{00}^2 Z_{n_z0}^2$ has a similar expression to the one for case
{\bf I)} above, but with the contribution $X_{00}=\int dx \,
\psi_{n_x=0}^2(x-\frac {a}{2})(\omega_{0}^2(x)+\omega_{0}(x)
\omega_{1}(x))$ , calculated using a shifted ground state wave
function.
\begin{figure}[t]
\centering
\includegraphics[scale=0.5]{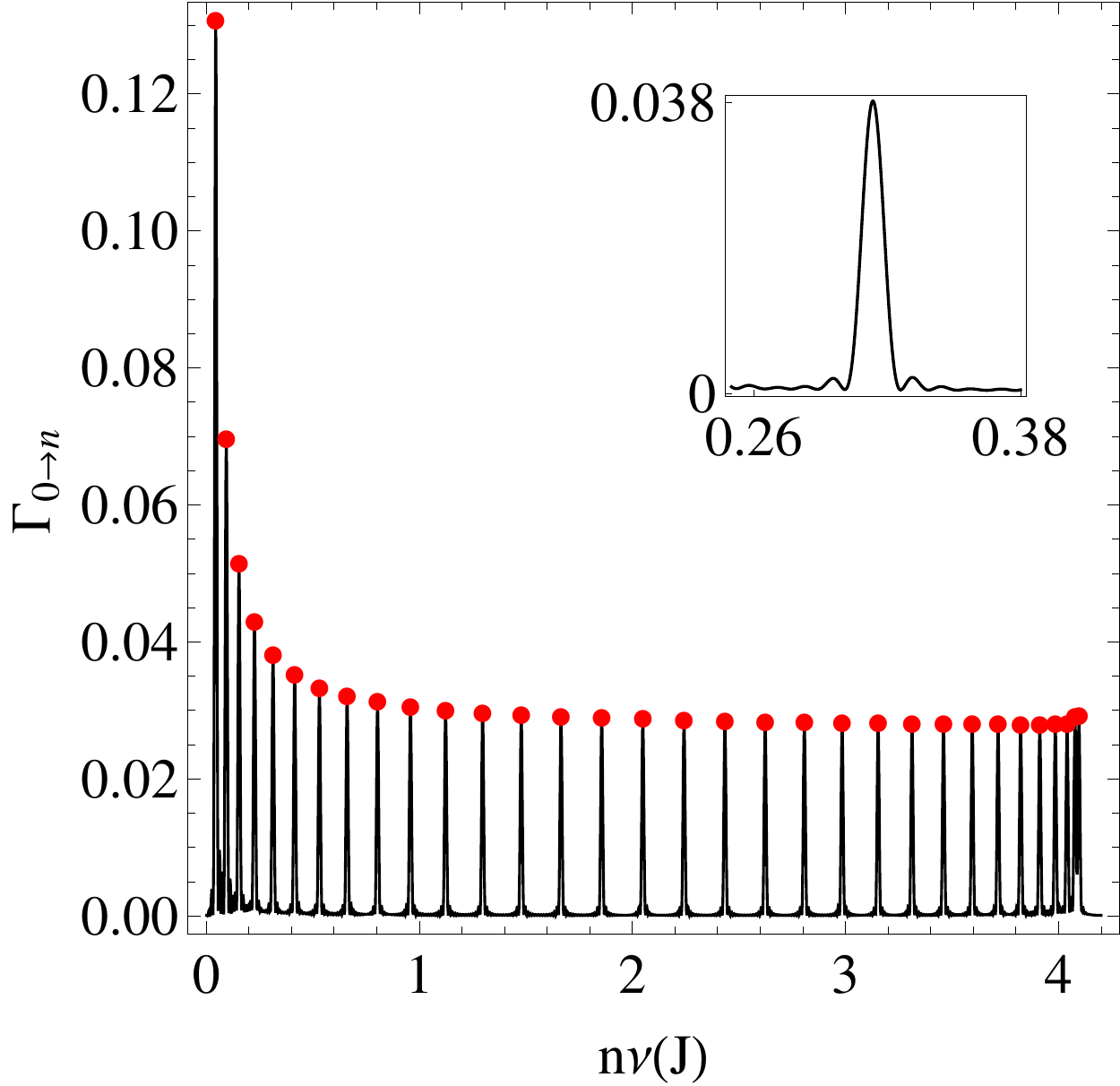}
\caption{Transition probabilities $\Gamma_{\bar 0
\rightarrow (0,0,n_z) }^{\textrm{\bf I}}$ for  {\bf \textrm{\bf I})}, as a
function of the probe energy gap for a fixed final time $g_{\textrm{\bf I},n}T_f \simeq 10^{-2}$. The
number of lattice sites is $N_{s}=65$, the temperature is
$\beta^{-1} = 1 \mbox{nK}$ and $J/U=10$, so that the lattice
bosons are in the superfluid regime. The red dots identify the
Bogoliubov frequencies. Inset: zoomed view of a transition peak. }
\label{spectrum1}
\end{figure}
These probabilities can be obtained experimentally by
i) initialising the probe in its ground state, and ii) measuring
the population of a selected excited state after a given time.
To reconstruct the excitation spectrum of the atomic gas, this
procedure should be repeated for different values of the energy
difference between the two involved impurity levels. This can be
done, in the harmonic case, by manipulating the frequency of the
probe confinement trap. Indeed, the transition probability
$\Gamma_{0\rightarrow n} $ is a function of the energy difference
between the probe levels as well as the overlap between the
lattice Wannier states and the unperturbed eigenfunctions of the
impurity. 
If the interaction time $T_f$ is large enough, resonance peaks
will emerge when scanning the probability $\Gamma_{0\rightarrow n}
$ for different trapping frequencies. Indeed, (for case {\bf I})
we have that $ \Gamma_{\bar 0 \rightarrow (0,0,n_z) }^{\textrm{\bf
I}}\simeq 2\;{g_{\textrm{\bf I},n}^{2}}\;\beta_k^2 n(\omega(k))
T_f^2  $. This probability is displayed in Fig.~\ref{spectrum1} as
a function of the impurity energy gap for a 65-site lattice at 1
nK. The peaks are located precisely at the frequencies of the
phononic excitations and their height is proportional to both the
occupation of each Bogoliubov mode and the spectral density
$\beta_k^2$; in particular, the progressive damping at higher
frequencies is also due to the thermal character of the atomic gas.
\begin{figure}[!t]
\centering
\includegraphics[scale=0.75]{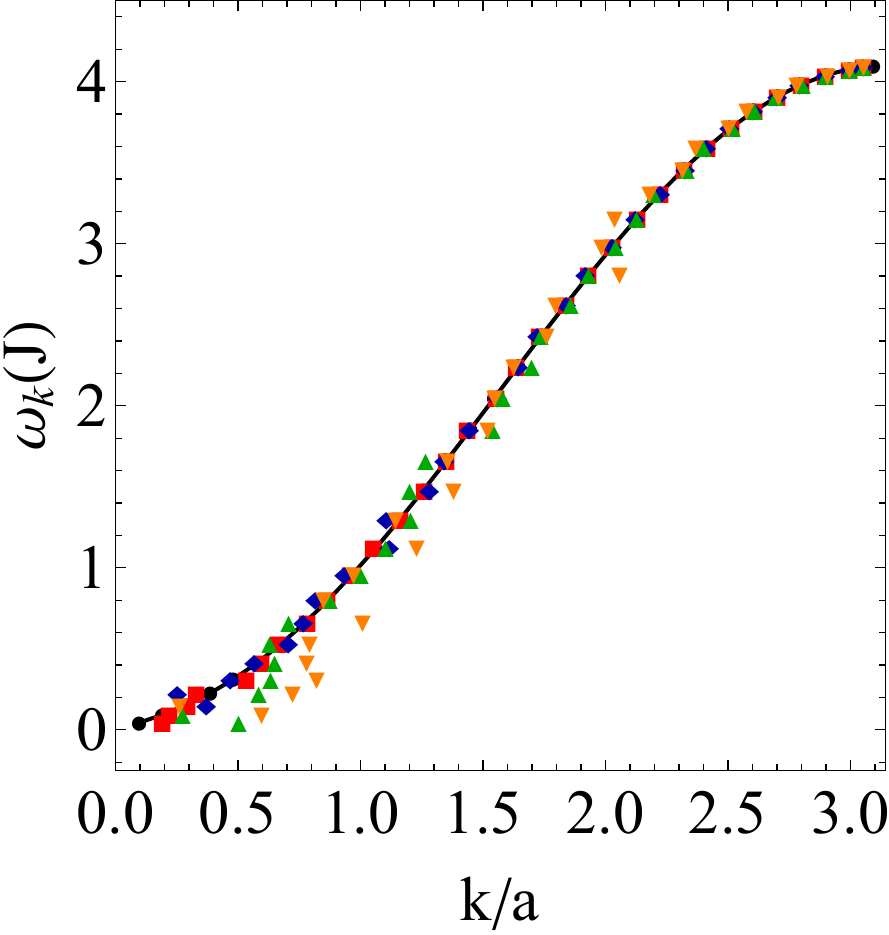}
\caption{ Comparison between the analytic excitation
spectrum $\omega(k)$ (black)
and the frequencies extracted via the probing protocol using the local
atomic probe in both configurations {\bf I} and {\bf II} described
in the text. The lattice parameters are the same as in
Fig.~\ref{spectrum1}.  Red, blue, green and orange markers correspond to 
a noise of f $1\%$, $2\%$, $5\%$ and
$10\%$ respectively.} \label{disprel1}
\end{figure}

\begin{figure}[!t]
\centering
\includegraphics[scale=0.75]{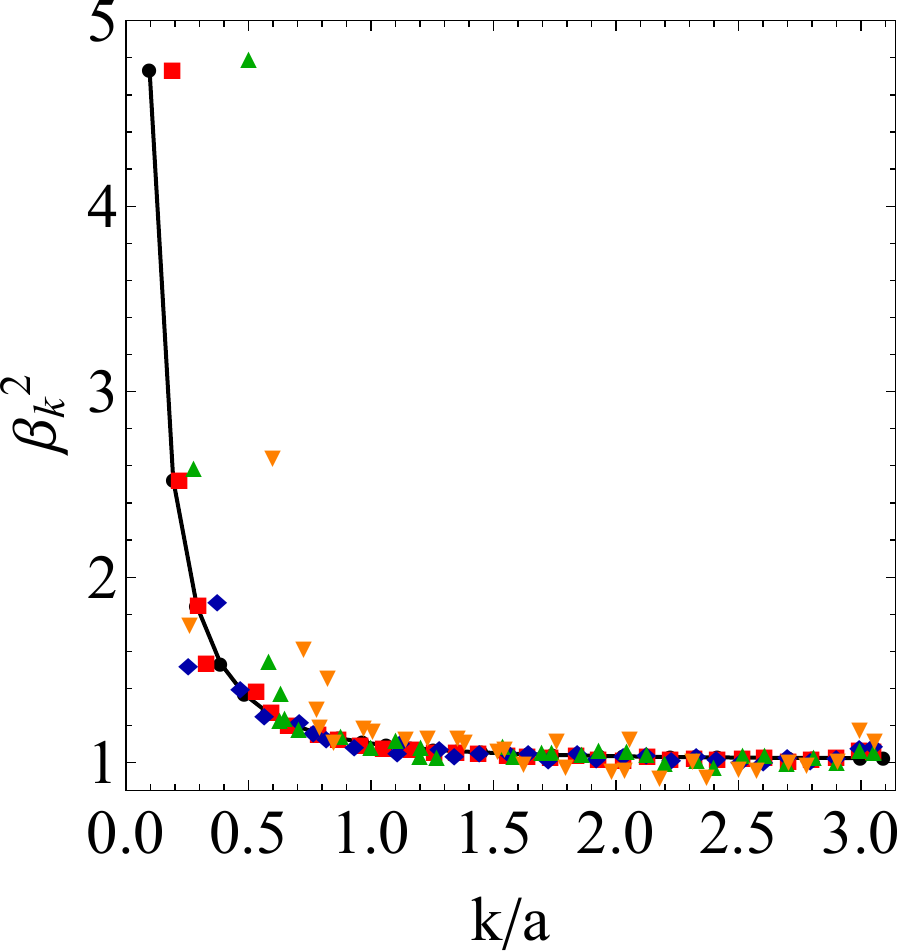}
\caption{Comparison between exact spectral function
$\beta_k^2$ (black line) and values extracted from
the rates with noise. The lattice parameters are the
same as in Fig.~\ref{spectrum1}. The legend is the same as in Fig.~\ref{disprel1}. } \label{betak}
\end{figure}
In order to
fully reconstruct the dispersion relation $\omega(k)$ and spectral density $\beta (k)$, their
dependence on the wave number $k$ is also needed, which can be
obtained via step {\bf II)}. When the
probe is located at a maximum of the optical potential and under
the assumption that its longitudinal confinement length $x_0$ is
comparable to the lattice constant $a$, its wave function overlaps
with the Wannier states of the two adjacent sites. This gives rise
to the extra $\cos (ka)$ factor in Eq. (\ref{gammaB}), which is
crucial in order to associate  the wave number $k$ to each
excitation frequency $\omega(k)$.  Again for a sufficiently long
interaction time, the transition probabilities simplifies
$
\Gamma_{\bar 0 \rightarrow (0,0,n_z) }^{\textrm{\bf II}}\simeq 2\;g_{\textrm{\bf II},n}^2 \;\beta_k^2 n(\omega(k)) T_f^2 (1+\cos (ka)),
$
leading to the following ratio
\begin{equation}
\frac { \Gamma^{\textrm{\bf II}}}{\Gamma^{\textrm{\bf I}}}\frac{g^{\textrm{\bf I}}}{g^{\textrm{\bf II}}}=\left[1+\cos (ka)\right].
\label{ratiogg}
\end{equation}
Therefore, by measuring both $\Gamma^{\textrm{\bf I}}$ and
$\Gamma^{\textrm{\bf II}}$, it is possible to discriminate the
wave number corresponding to each peak, thus probing the
Bogoliubov dispersion relation, even if the exact values of the
effective coupling constants are unknown. In other words, by
performing twice energy-resolved measurements of the impurity for
two different spatial configurations, it is possible to perform a
full momentum-resolved spectroscopy of the gas. This feature is
very important, since possible hopping between next nearest
neighbours will generate spectral features  resulting in a non
monotonic dispersion law that can only be captured when one can
discriminate energies in $k$. 
 {The reconstructed dispersion
relation is displayed in Fig.~\ref{disprel1}, in
comparison with the exact analytic values $\omega(k)$, (black line and dots). The reconstructed points are obtained from transition rates to whom a statistical noise of $1\%$, $2\%$, $5\%$ and
$10\%$ is applied, and a discrete sampling of the transition frequency is also taken into account. All the reconstructed curves are
able to capture the relevant features and behaviour of the the Bogoliubov spectrum. 
 One
can also extract the spectral density of Eq.~\ref{interaction2}.
Indeed, Fig.~\ref{betak} shows the comparison between the exact
spectral density (black line and dots) and the reconstructed ones 
for the same optical lattice and sources of error considered in Fig.~\ref{spectrum1} and  Fig.~\ref{disprel1}. The reconstructed relation is
able to capture the main features of  the exact spectral density.
We notice  that the low $k$ sector is more sensitive to noise, this is quite reasonable given
the nature of relation Eq.~\ref{ratiogg}. Low $k$ quantities are still visible in Fig.~\ref{spectrum1}, however is not possible to associate a proper wave vector. } The number of
excitations in a particular Bogoliubov mode strictly depends on
the temperature of the atomic gas. At low temperatures, high
energy excitations are mostly suppressed; therefore, a good
probing requires a larger interaction time. In this case, the
effective interaction strength appearing in the transition
probability becomes proportional to the trap frequency, {\it i.e.}
$ g \sim \nu$. As a result, all of the relevant parameters must be
chosen consistently with the perturbative approach. In particular,
to avoid coupling of the probe with bosons on more than two sites,
a crucial condition to fulfil is $m\nu_0>4/a^{2}$ (see
appendices for details). Even after this restriction
has been taken into account our probing scheme remains feasible
with current technology \cite{exp1,exp2,ott,osp}, considering specifically techniques to properly manipulate the impurity, checking its location and perform
the energy measurements \cite{expref2,expref3,expref4,expref5}. Furthermore, it can be
also applied to the Mott phase although only energy differences in
the Bogoliubov spectrum can be efficiently extracted in this case.
\section{Conclusions}
Concluding, we have presented an experimentally feasible protocol
to perform momentum-resolved spectroscopy of a 1D superfluid cold
gas in an optical lattice via energy-resolved measurements on
a single and controllable quantum system.
Our proposal exemplifies the essence of the quantum probing
approach, wherein some of the typical complexity of a many-body
system is broken down by imprinting it onto the open dynamics of a smaller system, and therefore locally extracted.
This process is not at all obvious  {\it a priori}.
Importantly, the protocol is potentially
non-invasive as it acts on the gas as a small density
perturbation, whose effects are rapidly suppressed after each
measurement. Furthermore, it can be extended to investigate other
lattice models, and generalised to a multi-probe schemes aimed at
studying genuine many-body features, such as
 quantum correlations.\\
\section*{Acknowledgements}
FC and SM acknowledge financial support from the Horizon 2020 EU collaborative project QuProCS (Grant Agreement 641277), the Academy of Finland Centre of Excellence program (Project no. 312058) and the Academy of Finland (Project no. 287750)

\appendix

\section{Details on the derivation of the transition rates}

The most general density-density interaction allowing for
transitions among different Wannier states takes the following
form
\begin{equation}
\begin{split}
\hat H_{int}=g\sum_{n,m;i,j}  \int dxdydz \; \psi_n^*(x,y,z) \psi_m(x,y,z) \omega_i(x)\omega_j(x)\\ \times \ket{n}\bra{m}\otimes \delta_{i}(y,z)\delta_{j}(y,z)  \hat c^\dagger_i \hat c_j,
\end{split}
\label{interaction2app}
\end{equation}
To compute the transition rates in the weak coupling regime one has to calculate the quantity $\bra {0,0,n_z} \hat H_{int}\ket{0,0,0}$ up to the first order in $g$. For the two cases considered in the manuscript we find\\
Case \textbf I
\begin{equation}
\begin{split}
\bra {0,0,n_z} \hat H_{int}&\ket{0,0,0} \simeq g \psi_{n_y=0}(y=0)^2 \psi_{n_z=0}(z=0)\psi_{n_z}(z=0) \\
&\times\int dx \,  \psi_{n_x=0}(x)^2\omega_0(x) \omega_0(x) \hat c^\dagger_0 \hat c_0
\end{split}
\label{averageA}
\end{equation}
Case \textbf II
\begin{equation}
\begin{aligned}
\bra {0,0,n_z} \hat H_{int}&\ket{0,0,0}
\simeq  g \psi_{n_y=0}(y=0)^2 \psi_{n_z=0}(z=0)\psi_{n_z}(z=0)\\
& \times\int dx \,  \psi_{n_x=0}(x-\frac{a}{2})^2\sum_{i,j=0,1} \omega_i(x)\omega_j(x) \hat c^\dagger_i \hat c_j
\label{averageB}
\end{aligned}
\end{equation}
In case \textbf I, the site in which the impurity is immersed is
labelled "$\mathbf 0$" and it is assumed that the interaction depends on the local
boson number operator related to this site only. In the next section we are going to
show that this local approximation is indeed good, calculating numerically the overlapping integrals.

In case \textbf {II}, instead, the impurity is embedded between sites
"$\mathbf 0$" and "$\mathbf 1$", and the terms that dominate the
dynamics are those containing the number operators of the two
sites as well as the transition operator describing tunnelling
between them. We then, need to expand the following combinations $
\hat c^\dagger_0 \hat c_0$, $ \hat c^\dagger_1 \hat c_1$, $ \hat
c^\dagger_0 \hat c_1$ and $ \hat c^\dagger_1 \hat c_0$ in terms of
the Bogoliubov operators of the condensate $\hat b_k = u_k \hat
c_k - v_k \hat c^\dagger_{-k}$, obtaining

\begin{equation}
\begin{split}
\hat c^\dagger_{i} \hat c_{j} = \left(\frac {1} {\sqrt {N_s}} \sum_k  e^{ikx_i}\hat c^\dagger_k\right)\left(\frac {1} {\sqrt {N_s}} \sum_q  e^{-iqx_{j}}\hat c_q\right)\\= \frac {1} {N_s}\left(\sqrt {N_0}+ \sum_{k\neq 0}  e^{ikx_i}\hat c^\dagger_k\right)\left(\sqrt {N_0}+ \sum_{q\neq 0}  e^{-iqx_{j}}\hat c_q\right)
\\\simeq n_0+ \sqrt {\frac{n_0}{N_s}}\sum_{k\neq 0} ( e^{ikx_i}\hat c^\dagger_k+ e^{-ikx_{j}}\hat c_k)
\\ =n_0+ \sqrt {\frac{n_0}{N_s}}\sum_{k\neq 0} \left[\left(u_k e^{ikx_i}+v_k e^{ikx_{j}})\hat b^\dagger_k+ (u_k e^{-ikx_{j}}+v_k e^{-ikx_{i}}\right)\hat b_k\right],
\end{split}
\end{equation}
in which, due to the linear nature of the Bogliubov approximation, we have neglected any process involving more than one non-zero momentum operator.
Hence, Eqs.~\ref{averageA}-\ref{averageB} can be recast as follows
\begin{equation}
\begin{split}
\bra {0,0,n_z} \hat H_{int} \ket{0,0,0}_a \simeq \phi
Y_{00}Z_{n_z0} \times  \nonumber \\  \left[n_0+ \sqrt
{n_0}\sum_{k\neq 0} \beta_k\left( e^{ikx_i}\hat b^\dagger_k+
e^{-ikx_i}\hat b_k\right)\right],
\label{averageA1}
\\
\bra{0,0,n_z} \hat H_{int}\ket{0,0,0}_b \simeq   Y_{00}Z_{n_z0} \times \nonumber \\
\Bigg\{2(\varphi + \varphi')n_0 +(\varphi+\varphi')\sqrt
{n_0}\sum_{k\neq 0} \beta_k\left[ \left(e^{ikx_{1}} +
e^{ikx_{0}}\right)\hat b^\dagger_k+ h.c. \right]\Bigg\} 
\end{split}
\end{equation}
in which we have defined the amplitude  $\beta_k =\sqrt
{\frac{n_0}{N_s}}(u_k+v_k)$, $\phi=\int dx \,
\psi_{n_x=0}^2(x)\omega_{0}(x) \omega_{0}(x)$, $\varphi=\int dx \,
\psi_{n_x=0}(x-a/2)^2\omega_{0/1}(x) \omega_{0/1}(x)$ and
$\varphi'=\int dx \,  \psi_{n_x=0}(x-a/2)^2\omega_{0/1}(x)
\omega_{1/0}(x)$.

\section{Physical Setting}
The boson condensate is trapped in a one-dimensional optical
potential $V(x)= V_0\sin^2 (\frac {2\pi x}{\lambda})$ that can be
generated using two laser beams, and having lattice parameter $a
=\frac {\lambda}{2}$. Different lattice potentials generate
Hubbard models with different $J$, $U$ and Wannier states. In
Fig.1 we show the values of the overlapping integrals for
different optical potentials as a function of the harmonic length
of the trapped impurity.
\begin{figure*}[t!]
\centering
\includegraphics[width=0.4\textwidth]{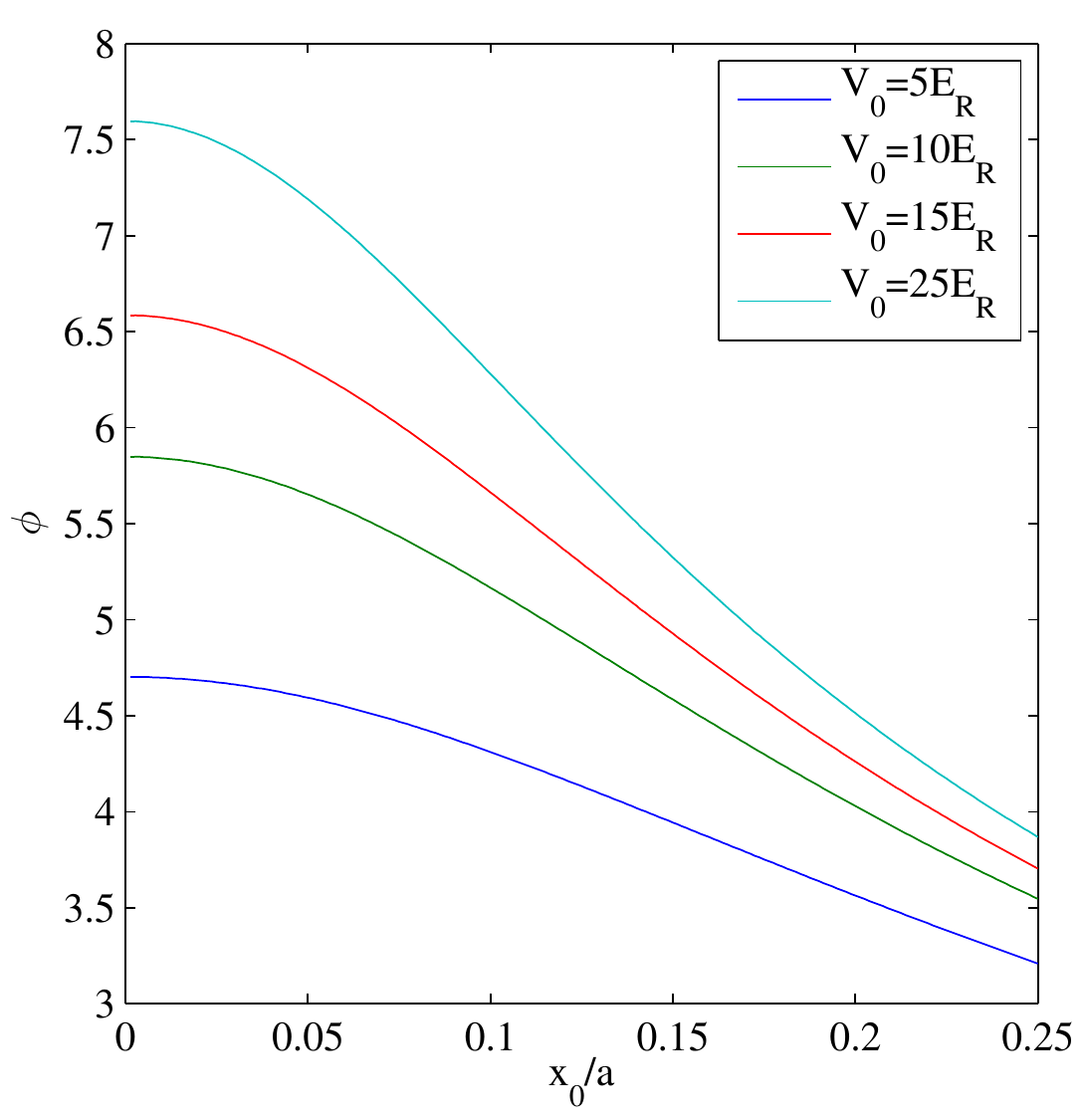}~~~\includegraphics[width=0.4\textwidth]{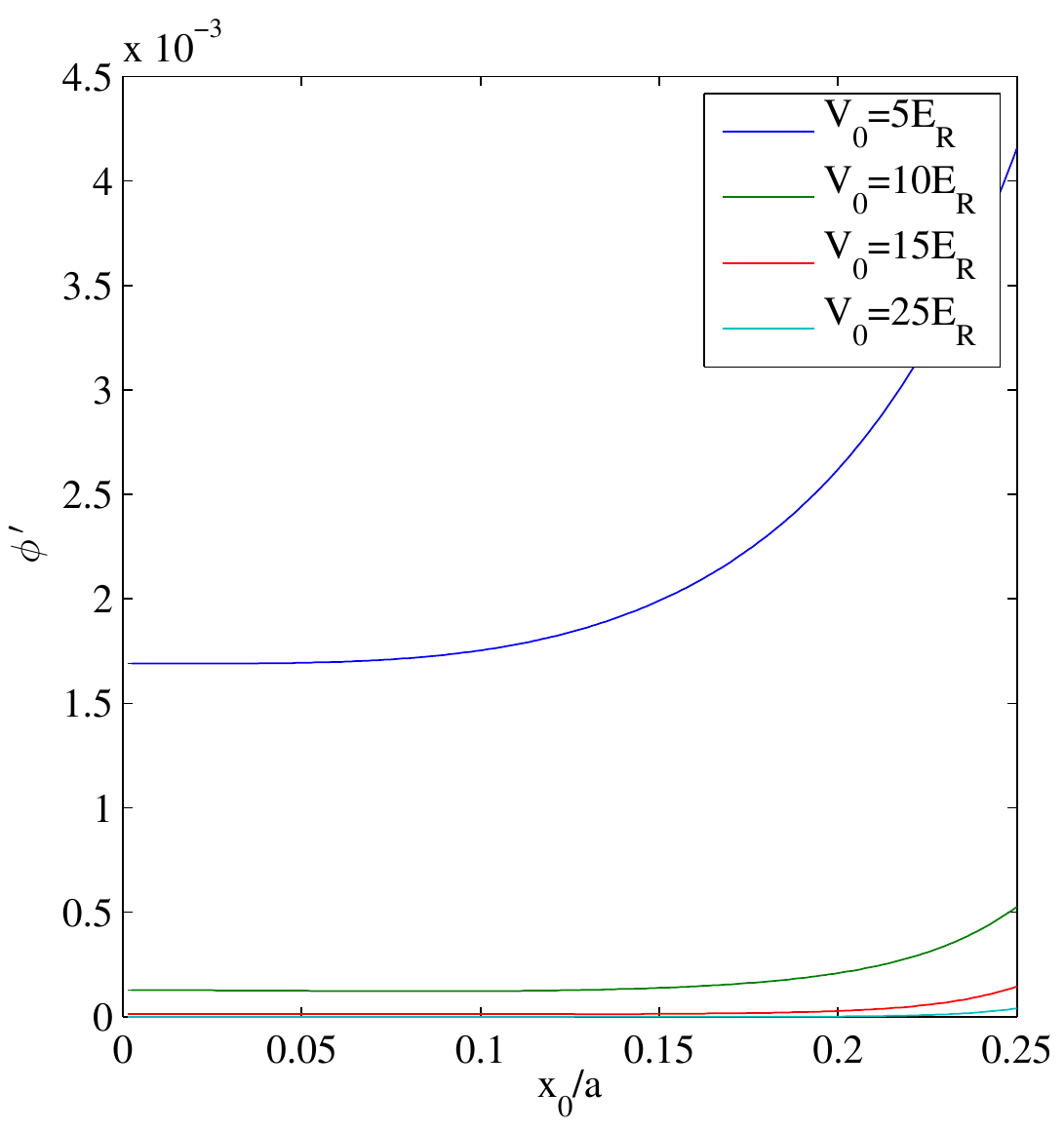}
\includegraphics[width=0.4\textwidth]{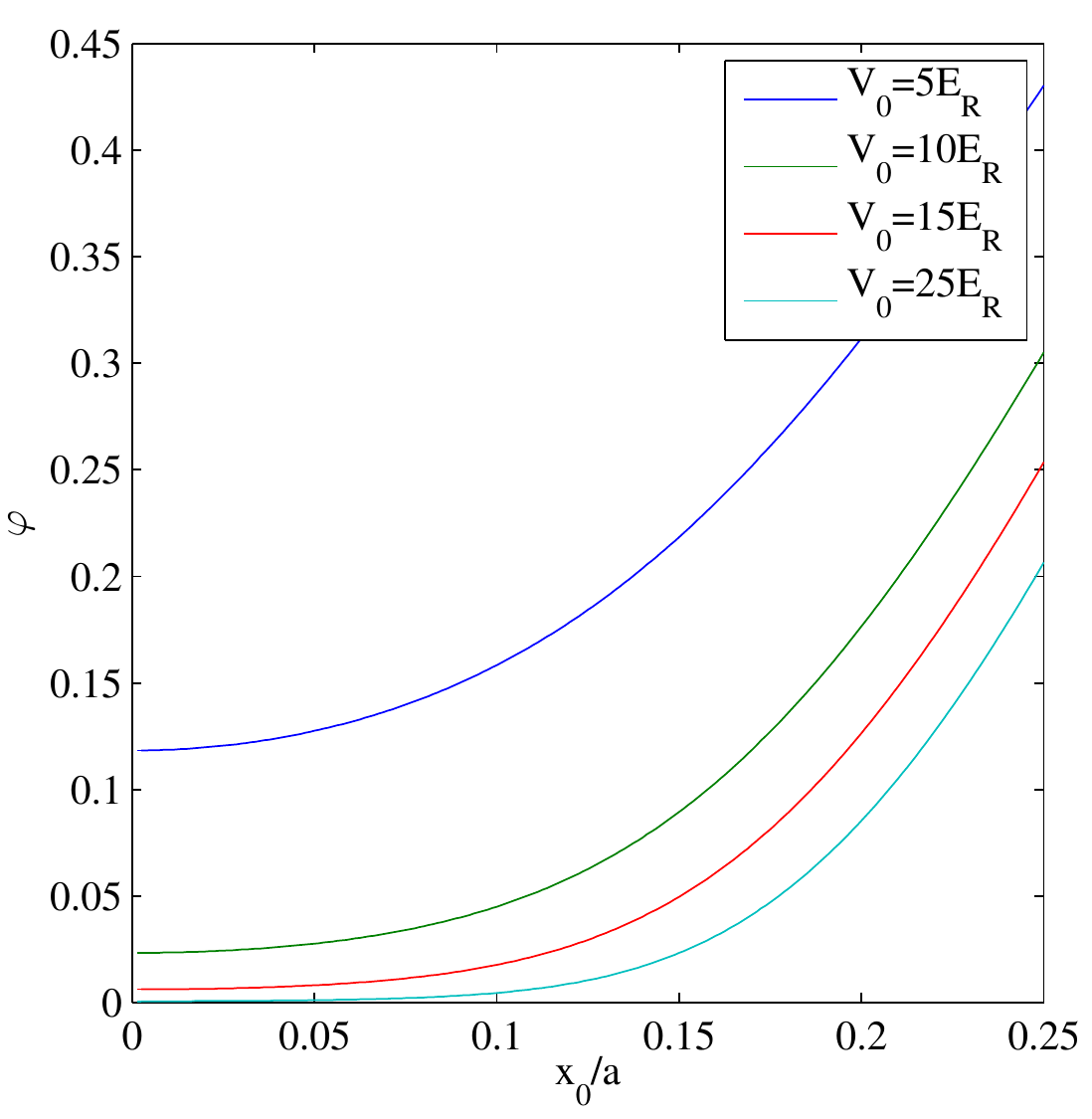}~~~\includegraphics[width=0.4\textwidth]{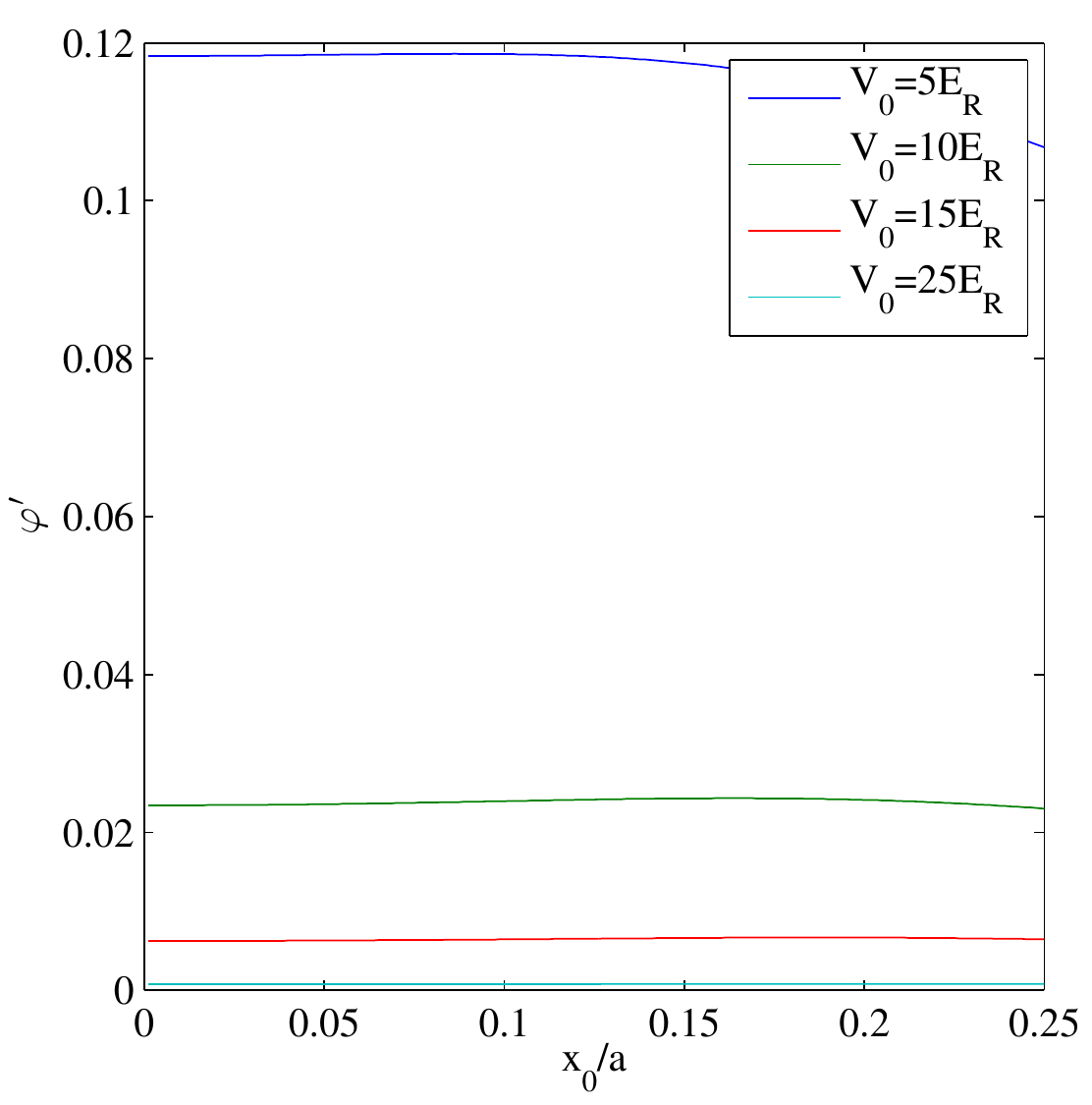}
\label{overlaps}
\caption {Upper panel: overlapping integrals $\phi$ (left) and $\phi'=\int dx\psi_{n_{x}=0}(x)^{2}\omega_{1}(x)^{2}$ (right). Lower panel: $\varphi$ (left) and $\varphi'$ (right). The overlapping integrals are computed using Wannier states for different potential depth expressed in terms of the recoil energy $E_{R}$.  Wannier states for different realizations of optical lattices have been generated using the software package developed at University of Oxford by the group of Prof. Dieter Jaksch \cite {w0}. The algorithm implemented in the software package is described in \cite {walters1}. For further details, see \cite {w2,w3}.}
\end{figure*}

Beside the expression given in the main text, the wave functions
of the impurity can be expressed in terms of the length $x_0=\sqrt
{\frac {\hbar}{m\nu}}$, $\psi_n(x)=\frac
{x_0^{-1/2}\pi^{-1/4}}{2^{n/2}n!^{1/2}}H_n(\frac {x}{x_0})
e^{-\frac {x^2}{2x_0^2} }$. For the case \textbf I we show $\phi$
and $\phi'$, they represent the integral containing the ground
state function of the probe and the Wannier function of the same
site and of its nearest neighbor respectively. We see how the
second one is far smaller than the first one. This proves that the
local approximation is good enough for an harmonically trapped
probe, within the specified trapping-length range.

\clearpage
\bibliographystyle{apsrev4-1}

\begin{thebibliography}{99}
\bibitem{nota} The same parity effect has been reported in Ref. \cite{plastina1}
\end{thebibliography}

\end{document}